\documentclass[a4paper,epsfig]{jpconf}

\usepackage{amssymb}
\usepackage{psfrag}

\usepackage{epsfig}
\usepackage{epsf}

\usepackage{amsmath}
\usepackage{amsfonts}
\usepackage{psfrag,epsfig,graphicx}

\usepackage{graphicx}

\newcommand{\pom}{\mathbb{P}}

\begin{document}
\title{Diffractive production of jets at high-energy in the QCD shock-wave approach}

\author{R Boussarie$^1$, A V Grabovsky $^{1,2,3}$, L Szymanowski$^{1,4,5}$ and S Wallon$^{1,6}$}

\address{$^1$ Laboratoire de Physique Th\'{e}orique, CNRS, Universit\'{e} Paris Sud, Universit\'{e} Paris Saclay, \hspace*{0.3cm}91405 Orsay, France}
\address{$^2$ 
Novosibirsk State University,
2 Pirogova street, Novosibirsk, Russia}
\address{$^3$ 
Theory division, Budker Institute of Nuclear Physics,
11 Lavrenteva avenue, Novosibirsk, \hspace*{0.3cm}Russia}
\address{$^4$ National Centre for Nuclear Research, Warsaw, Poland}
\address{$^5$ Centre de Physique Th\'eorique, Ecole Polytechnique, CNRS, Universit\'e Paris-Saclay, F91128 \hspace*{0.3cm}Palaiseau, France}
\address{$^6$ UPMC Universit\'{e} Paris 6, Facult\'{e} de physique, 4 place Jussieu, 75252 Paris Cedex 05, France}

\ead{renaud.boussarie@th.u-psud.fr, a.v.grabovsky@inp.nsk.su, lech.szymanowski@ncbj.gov.pl and samuel.wallon@th.u-psud.fr}

\begin{abstract}	
In view of future phenomenological applications, we study the impact factor for the photon to quark, antiquark and gluon transition within
Balitsky's shock-wave formalism. 
The aim of the present program is to extend existing results
beyond approximations discussed in the literature.
We present our results of the real contribution, and report on recent progress in calculating the virtual contributions for the photon to quark, antiquark transition. 

\end{abstract}

\section{Introduction}

\begin{figure}[h]
\hspace{0cm}\includegraphics[scale=0.28]{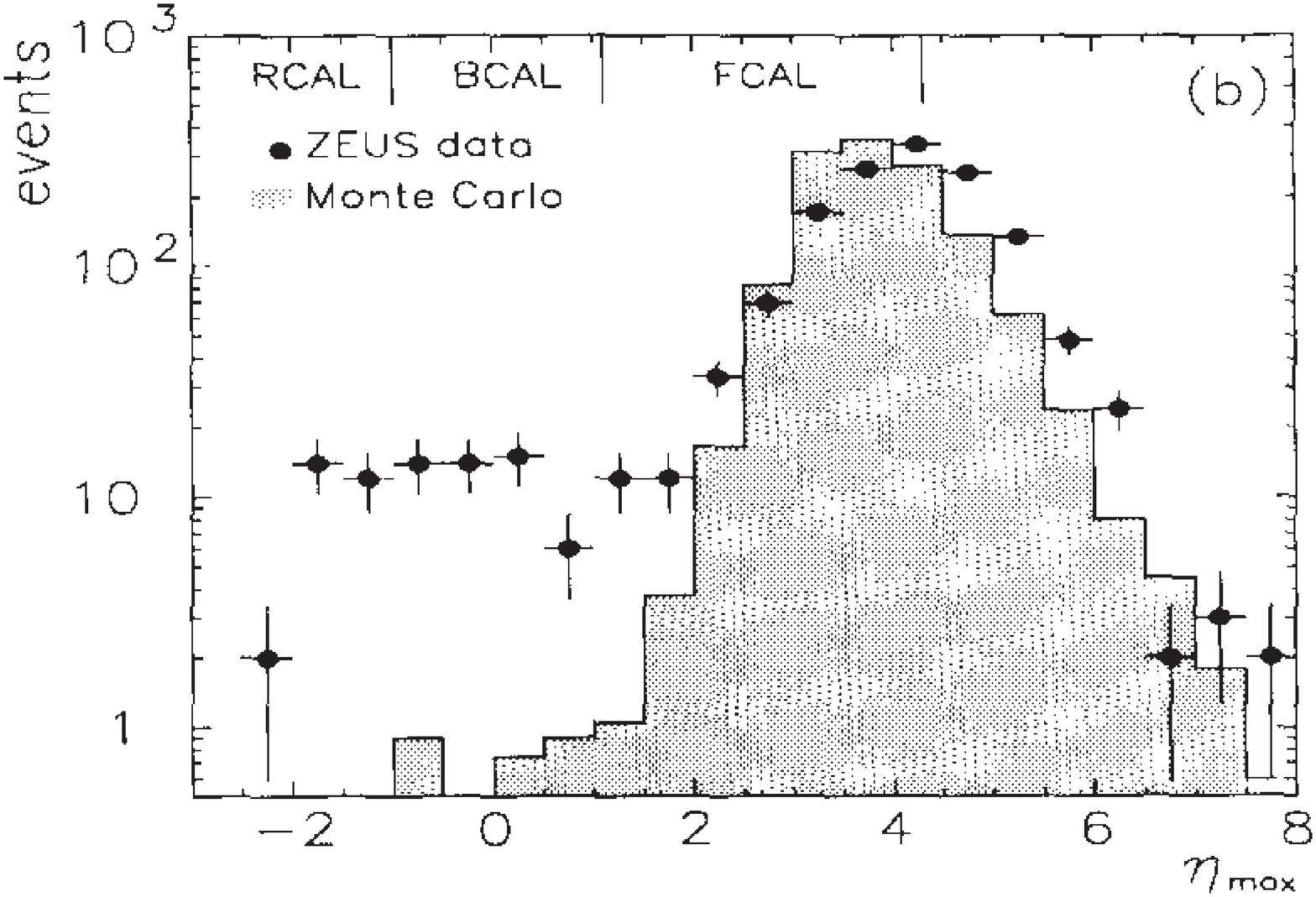}
\qquad
\hspace{0.2cm}\includegraphics[scale=0.28]{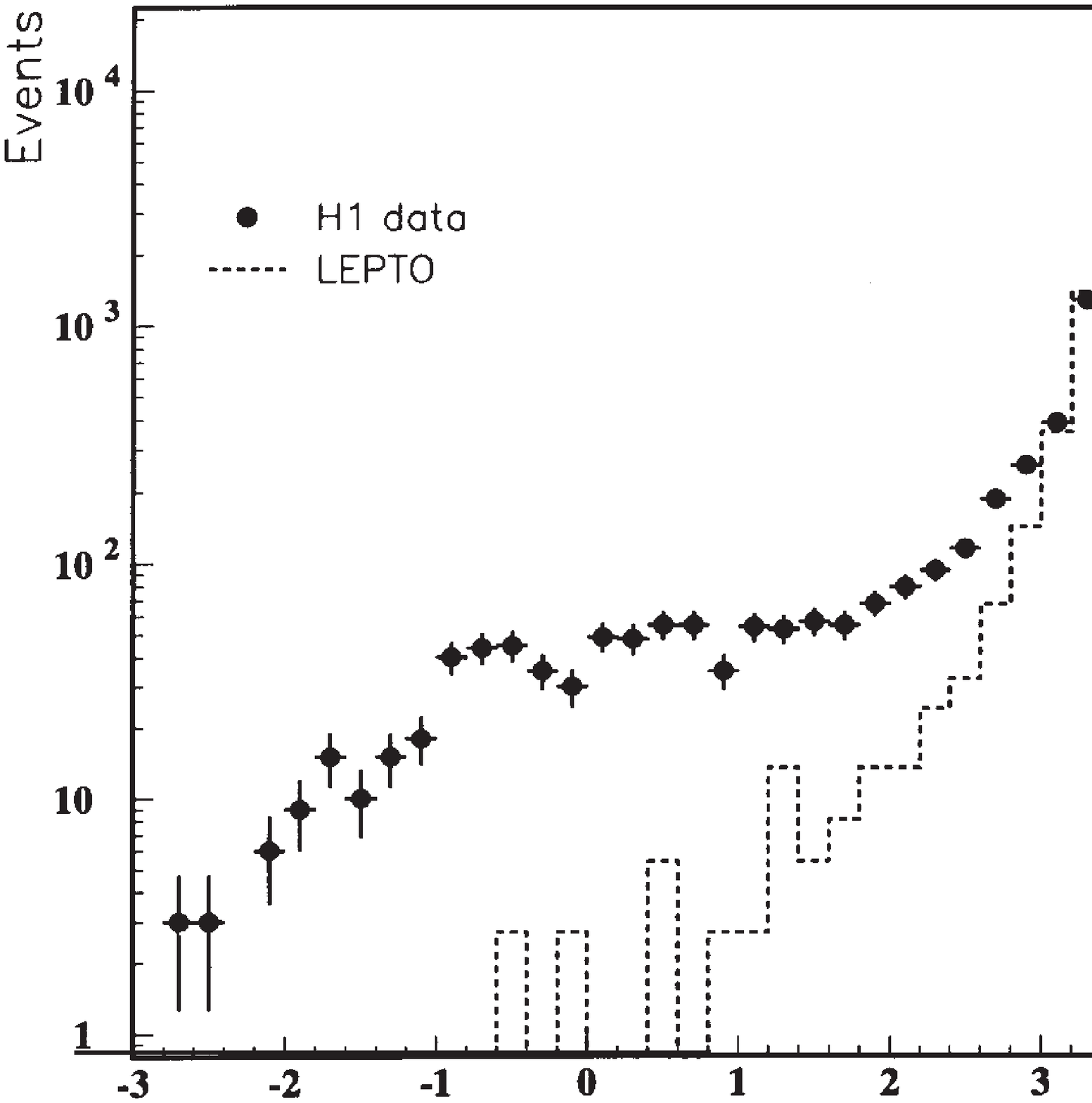}
\caption{Excess of events exhibiting a rapidity gap with the proton remnants with respect to Monte Carlo DIS predictions, providing an evidence for diffractive events, at ZEUS (1993) (left panel)~\cite{Derrick:1993xh} and H1 (1994) (right panel)~\cite{Ahmed:1994nw}.}
\label{ZEUS-H1}
\end{figure}

One of the major achievements of HERA was the experimental evidence \cite{Derrick:1993xh,Ahmed:1994nw}
shown in Fig.~\ref{ZEUS-H1} that 
among the whole set of $\gamma^* p \to X$ deep inelastic scattering events, almost 10\%  are diffractive (DDIS), of the form $\gamma^* p \to X Y$ with a rapidity gap between the proton remnants $Y$
and the hadrons $X$
coming from the fragmentation region of the initial virtual photon. 

\begin{figure}[h]
\center
\psfrag{q}{\raisebox{-.2cm}{$\gamma^*$}}
\psfrag{l1}{$e^\pm$}
\psfrag{l2}{$e^\pm$}
\psfrag{P}{$\pom$}
\psfrag{ld}{}
\psfrag{lu}{}
\psfrag{R}{}
\psfrag{q1}{\raisebox{.2cm}{\ \ jet}}
\psfrag{q2}{\raisebox{-.3cm}{\ \ jet}}
\psfrag{p1}{$p$}
\psfrag{p2}{$Y$}
\includegraphics[scale=.90]{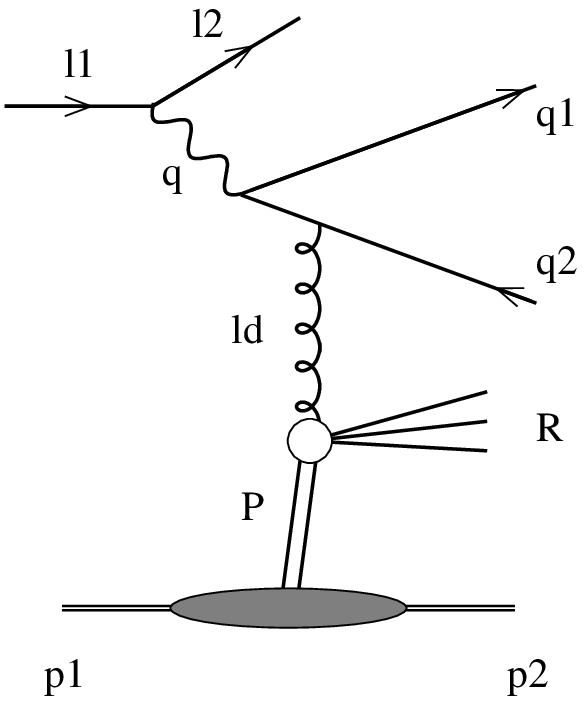}
\qquad 
\psfrag{q2}{\raisebox{-.4cm}{\ \ jet}}
\raisebox{.5cm}{\includegraphics[scale=.90]{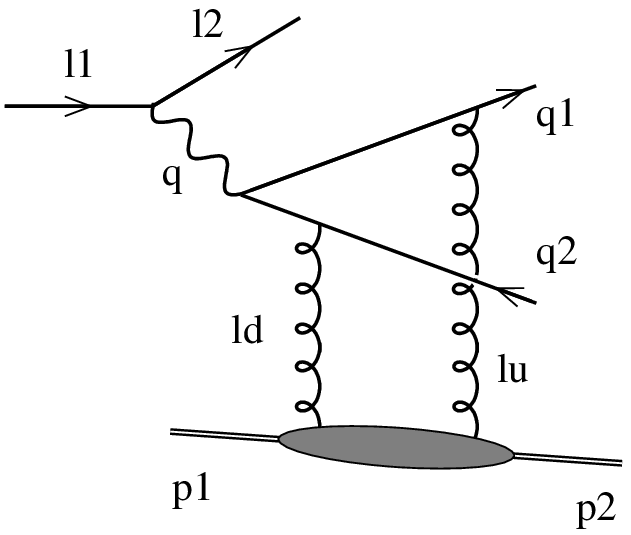}}
\caption{Resolved (left panel) and direct Pomeron (right panel) contributions to two jets production.}
\label{ResDirect}
\end{figure}

Diffraction can be theoretically described according to several approaches. The first one involves
a {\em resolved} Pomeron contribution (with a parton distribution function inside the Pomeron), see Fig.~\ref{ResDirect} (left),  while the second one
relies on a  {\em direct} Pomeron contribution involving the coupling of a Pomeron with the diffractive state, see Fig.~\ref{ResDirect} (right). 

The diffractive states can be modelled in perturbation theory by a  $q \bar{q}$ pair (for moderate $M^2$, where $M$ is the invariant mass of the  diffractively produced state $X$) or by higher Fock states as a $q \bar{q} g$ state for larger values of $M^2$. Based on such a model, with
 a two-gluon exchange picture for the Pomeron,  a good description of HERA data for diffraction~\cite{Chekanov:2004hy-Chekanov:2005vv-Chekanov:2008fh,
Aktas:2006hx-Aktas:2006hy-Aaron:2010aa-Aaron:2012ad-Aaron:2012hua} could be achieved~\cite{Bartels:1998ea}. One of the important features of this approach is that the $q \bar{q}$ component with a longitudinally polarized photon plays a crucial role in the region of small
diffractive mass $M$, although it is a
twist-4 contribution.
In the direct components considered there, the $q \bar{q} g$ diffractive state has been studied in two particular limits. The first one, valid for very large $Q^2$, corresponds to a collinear approximation in which the transverse momentum of the gluon is assumed to be much smaller than the transverse momentum of the emitter~\cite{Wusthoff:1995hd-Wusthoff:1997fz}. 
The second one~\cite{Bartels:1999tn,Bartels:2002ri}, valid for very large $M^2$, is based on the assumption of a strong ordering of longitudinal momenta, encountered in BFKL equation~\cite{Fadin:1975cb-Kuraev:1976ge-Kuraev:1977fs-Balitsky:1978ic}. Both these approaches were combined in order to describe HERA data for DDIS~\cite{Marquet:2007nf}. 

Based on these very successful developments led at HERA
in order to understand the QCD dynamics with diffractive events, 
it would be appropriate to look for similar hard diffractive events at LHC. 
The idea there is to adapt the concept of  photoproduction of diffractive jets, which  was performed at HERA~\cite{Chekanov:2007rh,Aaron:2010su}, now with a flux of
quasi-real photons in ultraperipheral collisions (UPC)~\cite{Baltz:2007kq-Baur:2001jj}, relying on the notion of equivalent photon approximation. In both cases, 
 the hard scale is provided by the invariant mass of the tagged jets.

We here report on our computation~\cite{Boussarie:2014lxa} of the $\gamma^* \to q \bar{q} g$ impact factor at tree level with an arbitrary number of $t$-channel gluons described within the Wilson line formalism, also called QCD shockwave approach~\cite{Balitsky:1995ub-Balitsky:1998kc-Balitsky:1998ya-Balitsky:2001re}. As an aside, we rederive the $\gamma^* \to q \bar{q}$ impact factor. In particular, the 
$\gamma^* \to q \bar{q} g$ transition is computed without any soft or collinear approximation for the emitted gluon, in contrast with the above mentioned calculations. These results provide necessary generalization of building blocks for inclusive DDIS as well as for two- and three-jet diffractive production. Since the results we derived can account for an arbitrary number of $t$-channel gluons, this could allow to include higher twist effects which are suspected to be rather important in DDIS for $Q^2 \lesssim 5$ GeV$^2$~\cite{Motyka:2012ty}.

\section{An introduction to the shockwave formalism}

As stated before, we use Balitsky's shockwave formalism. 
Its application shows that this method is very powerful in determining evolution equations and impact factors at next-to-leading order for inclusive processes~\cite{Balitsky:2010ze-Balitsky:2012bs}, at semi-inclusive level for $p_t$-broadening in $pA$ collisions~\cite{Chirilli:2011km-Chirilli:2012jd} or in the evaluation of the triple Pomeron vertex beyond the planar limit~\cite{Chirilli:2010mw}, when compared with usual methods based on summation of contributions of individual Feynman diagrams computed in momentum space. It is an effective way of estimating the effect of multigluon exchange. Its formulation in coordinate space makes it natural in view of describing saturation~\cite{GolecBiernat:1998js-GolecBiernat:1999qd}. \\

We use the following notation. We introduce the light cone vectors
$n_{1}$ and $n_{2}$%
\begin{equation}
\label{Sudakov-basis}
n_{1}=\left(  1,0,0,1\right)  ,\quad n_{2}=\frac{1}{2}\left(  1,0,0,-1\right)
,\quad n_{1}^{+}=n_{2}^{-}=n_{1} \cdot n_{2}=1
\end{equation}
Then, 
one introduces Wilson lines as 
\begin{equation}
U_{i}=U_{\vec{z}_{i}}=U\left(  \vec{z}_{i},\eta\right)  =P \exp\left[{ig\int_{-\infty
}^{+\infty}b_{\eta}^{-}(z_{i}^{+},\vec{z}_{i}) \, dz_{i}^{+}}\right]\,.
\label{WL}%
\end{equation}
The operator $b_{\eta}^{-}$ is the external shock-wave field built from slow gluons 
whose momenta are limited by the longitudinal cut-off defined by the rapidity $\eta$
\begin{equation}
b_{\eta}^{-}=\int\frac{d^{4}p}{\left(  2\pi\right)  ^{4}}e^{-ip \cdot z}b^{-}\left(
p\right)  \theta\left(e^{\eta}-\frac{|p^{+}|}{P^+}\right)\,,\label{cutoff}%
\end{equation}
where $P^+$ is the typical large $+$ momentum of the problem, to be identified with $p_\gamma^+$ later on. We will denote the longitudinal cut-off $\sigma = e^\eta \, P^+ = \alpha \, P^+.$

We use the light cone gauge
$\mathcal{A}\cdot n_{2}=0,$
with $\mathcal{A}$ being the sum of the external field $b$ and the quantum field
$A$%
\begin{equation}
\mathcal{A}^{\mu} = A^{\mu}+b^{\mu},\;\;\;\;\;\;\;\;\;\quad b^{\mu}\left(  z\right)  =b^{-}(z^{+},\vec{z}\,) \,n_{2}%
^{\mu}=\delta(z^{+})B\left(  \vec{z}\,\right)  n_{2}^{\mu}\,,\label{b}%
\end{equation}
where
$B(\vec{z})$ is a profile function. \\
Indeed, let us consider an external gluon field $b^{\mu}$ in its rest frame and boost it along the $+$ direction. One obtains :
\begin{eqnarray}\nonumber
&&b^+ \! \left( x^+,\, x^-, \, \vec{x} \right) \rightarrow \frac{1}{\lambda}b^+ \left( \lambda x^+,\, \frac{1}{\lambda} x^- ,\, \vec{x} \right)\,, \\ \nonumber
&&b^- \! \left( x^+,\, x^-, \, \vec{x} \right) \rightarrow {\lambda} b^- \left( {\lambda x^+},\, {\frac{1}{\lambda} x^-} ,\, \vec{x} \right)\,, \\ \nonumber
&&b^i \, \left( x^+,\, x^-, \, \vec{x} \right) \rightarrow \, \, \, b^i \, \left( \lambda x^+,\, \frac{1}{\lambda} x^- ,\, \vec{x} \right)\,. \\ \nonumber
\end{eqnarray}
Assuming that the field vanishes at infinity, one immediately gets that only its minus component survives the boost in the limit $\lambda \to \infty\,,$ and that it does not depend on $x^-$ and contains $\delta \left( x^+ \right)\,,$ thus justifying the form of $b^\mu$ in Eq.~(\ref{b}).

\noindent
The natural and extensively used operator appearing in studies of diffractive processes within the shock wave approach is the dipole operator 
$\mathbf{U}_{12}=\frac{1}{N_{c}}\rm{tr}\left(  U_{1}U_{2}^{\dagger}\right)  -1$ constructed from the Wilson line (\ref{WL}).

\section{Impact factor for $\gamma\rightarrow q\bar{q}$ transition}

\begin{figure}
\center
\includegraphics[scale=0.65]{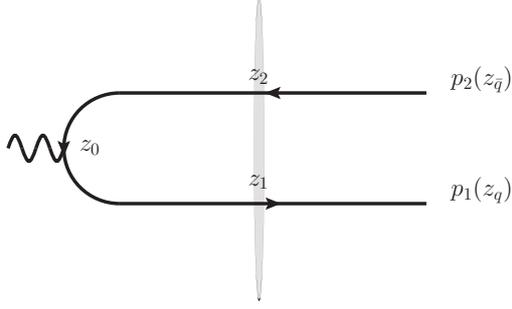}
\caption{Diagram contributing to the impact factor for two jet production }
\label{leading}
\end{figure}

\noindent
In the leading order the diagram contributing to the impact factor  for $\gamma\rightarrow q\bar{q}$ transition is shown in Fig. \ref{leading}, in which $z's$ denote the 
coordinates of interaction points with the photon and the shock wave.
After projection on the color singlet state and subtraction of the contribution without interaction with the shock wave, the contribution of this diagram can be written in the momentum space as (factorizing out a global QED factor $-i e_q$)
\begin{equation}
M_{0}^{\alpha}=N_c \int d\vec{z}_{1}d\vec{z}_{2}F\left(  p_{q},p_{\bar{q}}%
,z_{0},\vec{z}_{1},\vec{z}_{2}\right)  ^{\alpha} \mathbf{U}_{12}\,.
\label{M0int}%
\end{equation}
Denoting $Z_{12} = \sqrt{x_{q}x_{\bar{q}}\vec{z}_{12}^{\,\,2}}$, we get for a longitudinally polarized photon
\begin{eqnarray}
\label{FL}
F\left(  p_{q},p_{\bar{q}},k,\vec{z}_{1},\vec{z}_{2}\right)  ^{\alpha
}\varepsilon_{L\alpha}
\nonumber \\
&&\hspace*{-4cm}=\theta(p_{q}^{+})\,\theta(p_{\bar{q}}^{+})\frac
{\delta\left(  k^{+}-p_{q}^{+}-p_{\bar{q}}^{+}\right)  }{(2\pi)^{2}}%
e^{-i\vec{p}_{q}\cdot \vec{z}_{1}-i\vec{p}_{_{\bar{q}}}\cdot\vec{z}_{2}}
(-2i)\delta_{\lambda_{q},-\lambda_{\bar{q}}}\,x_{q}x_{\bar{q}}%
\,Q\,K_{0}\left(Q \, Z_{12}\right)\,,
\end{eqnarray}
whereas for a transversally polarized photon
\begin{eqnarray}
\label{FT}
F(  p_{q},p_{\bar{q}},k,\vec{z}_{1},\vec{z}_{2})  ^{j}%
\varepsilon_{Tj}\!
\nonumber \\
&&\hspace*{-5cm}=\theta(p_{q}^{+})\,\theta(p_{\bar{q}}^{+})\frac{\delta(
k^{+}\!\!-\!p_{q}^{+}\!-p_{\bar{q}}^{+}\!)  }{(2\pi)^{2}}e^{-i\vec{p}_{q}\cdot\vec
{z}_{1}-i\vec{p}_{_{\bar{q}}}\cdot\vec{z}_{2}}
\delta_{\lambda_{q},-\lambda_{\bar{q}}}( x_{q}-x_{\bar{q}%
}+s\lambda_{q})  \frac{\vec{z}_{12} \cdot \vec{\varepsilon}_{T}}{\vec{z}_{12}^{\,\,2}}
Q \,Z_{12} K_{1}(Q\, Z_{12})\,.\!\!\!\!\!
\end{eqnarray}

\section{Impact factor for $\gamma\rightarrow q\bar{q}g$ transition}

\begin{figure}
\center
\includegraphics[scale=0.65]{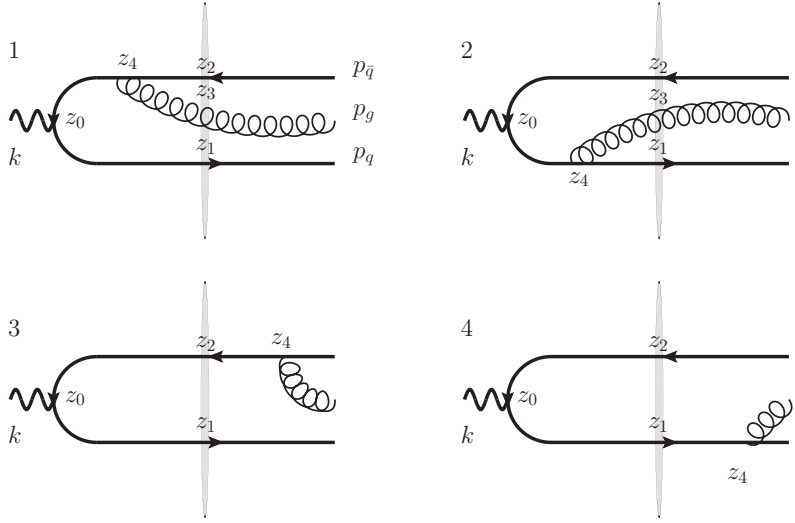}
\caption{Diagrams contributing to the impact factor for three jet production}
\label{3body}
\end{figure}

In the case of the $q\,\bar q\,g$ Fock final state the contributiong diagrams are shown in Fig.~\ref{3body}.
After projection on the color singlet state and subtraction of the contribution without interaction with the shock wave,  the result can be put in the form 
\begin{eqnarray}
\nonumber 
M^{\alpha} &=& N_c^2 \int d\vec{z}_{1}d\vec{z}_{2}d\vec{z}_{3} \, F_{1}\left(  p_{q},p_{\bar{q}}%
,p_{g},z_{0},\vec{z}_{1},\vec{z}_{2},\vec{z}_{3}\right)  ^{\alpha}\frac{1}{2}
\left( \mathbf{U}_{32} + \mathbf{U}_{13} - \mathbf{U}_{12} + \mathbf{U}_{32}\,\mathbf{U}_{13} \right)
\\
&+& N_c \int d\vec{z}_{1}d\vec{z}_{2} \, F_{2}\left(  p_{q},p_{\bar{q}},p_{g},z_{0}%
,\vec{z}_{1},\vec{z}_{2}\right)  ^{\alpha}\frac{N_{c}^{2}-1}{2N_{c}} \mathbf{U}_{12}\,.
\label{F2tilde}%
\end{eqnarray}
The first and the second line of this equation correspond to contributions to the impact factor, respectively, of the diagrams 1 and 2  of      Fig.~\ref{3body} and of 
the diagrams 3 and 4 of it.
For a longitudinally polarized photon, the functions $F_i$ read
\begin{eqnarray}
&&\hspace{-1cm}F_{1}\left(  p_{q},p_{\bar{q}},p_{g},k,\vec{z}_{1},\vec{z}_{2},\vec{z}%
_{3}\right)  ^{\alpha}\varepsilon_{L\alpha}=2\,Q\,g\,\delta(k^{+}-p_{g}^{+}-p_{q}%
^{+}-p_{_{\bar{q}}}^{+})\theta(p_{g}^{+}-\sigma)\frac{e^{-i\vec{p}_{q} \cdot %
\vec{z}_{1}-i\vec{p}_{_{\bar{q}}} \cdot \vec{z}_{_{2}}-i\vec{p}_{g} \cdot \vec{z}_{3}}}%
{\pi\sqrt{2p_{g}^{+}}}
\nonumber \\
&\times&\delta_{\lambda_{q},-\lambda_{\bar{q}}}\left\{  (x_{_{\bar{q}}}%
+x_{g}\delta_{-s_{g}\lambda_{q}})x_{q}\frac{\vec{z}_{32} \cdot \vec{\varepsilon}%
_{g}^{\,\,\ast}}{\vec{z}_{32}^{\,\,2}}-(x_{q}+x_{g}\delta_{-s_{g}%
\lambda_{\bar{q}}})x_{_{\bar{q}}}\frac{\vec{z}_{31} \cdot \vec{\varepsilon}%
_{g}^{\,\,\ast}}{\vec{z}_{31}^{\,\,2}}\right\} K_{0}(QZ_{123}) \,,\\
\label{F1eL}%
\label{resF2L}
&&\hspace{-1cm}\tilde{F}_{2}\left(  p_{q},p_{\bar{q}},p_{g},k,\vec{z}_{1},\vec{z}_{2}\right)
^{\alpha}\varepsilon_{L\alpha}=4ig \, Q\,\theta(p_{g}^{+}-\sigma)\delta(k^{+}%
-p_{g}^{+}-p_{q}^{+}-p_{_{\bar{q}}}^{+})\frac{e^{-i\vec{p}_{q} \cdot \vec{z}%
_{1}-i\vec{p}_{_{\bar{q}}} \cdot \vec{z}_{2}}}{\sqrt{2p_{g}^{+}}}%
\nonumber \\
&&\times\delta_{\lambda_{q},-\lambda_{\bar{q}}}\frac{x_{q}\left(  x_{g}%
+x_{\bar{q}}\right)  \left(  \delta_{-s_{g}\lambda_{q}}x_{g}+x_{\bar{q}%
}\right)  }{x_{\bar{q}} \, x_g }\frac{\vec{P}_{\bar{q}} \cdot %
\vec{\varepsilon}_{g}^{\,\,\ast}}{\vec{P}_{\bar{q}}^2}\,e^{-i\vec{p}_{g} \cdot \vec{z}_{2}}K_{0}%
(QZ_{122})-\left(  q\leftrightarrow\bar{q}\right)  ,
\end{eqnarray}
while for a transversally polarized photon, we have 
\begin{eqnarray}
&&\hspace{-.7cm}F_{1}\left(  p_{q},p_{\bar{q}},p_{g},k,\vec{z}_{1},\vec{z}_{2},\vec{z}%
_{3}\right)  ^{\alpha}\!\varepsilon_{T\alpha}=\!-2i\,g\,Q\delta(k^{+}-p_{g}^{+}%
-p_{q}^{+}-p_{_{\bar{q}}}^{+})\theta(p_{g}^{+}-\sigma)
\frac{e^{-i\vec{p}_{q} \cdot \vec{z}_{1}-i\vec{p}_{_{\bar{q}}} \cdot \vec{z}_{_{2}%
}-i\vec{p}_{g} \cdot \vec{z}_{3}}}{\pi Z_{123}\sqrt{2p_{g}^{+}}}
\nonumber \\
&&\delta_{\lambda_{q},-\lambda_{\bar{q}}}K_{1}(QZ_{123})\,\left\{  \frac{\left(  \vec{z}%
_{23} \cdot \vec{\varepsilon}_{g}^{\,\,\ast}\right)  \left(  \vec{z}_{13} \cdot %
\vec{\varepsilon}_{T}\right)  }{\vec{z}_{23}{}^{2}}x_{q}\left(  x_{q}%
-\delta_{s\lambda_{\bar{q}}}\right)  \left(  x_{\bar{q}} +x_{g}\delta
_{-s_{g}\lambda_{q}}\right)  \right.
\nonumber \\
&& \left.
+\frac{\left(  \vec{z}_{23} \cdot \vec{\varepsilon}_{g}^{\,\,\ast}\right)
\left(  \vec{z}_{23} \cdot \vec{\varepsilon}_{T}\right)  }{\vec{z}_{23}{}^{2}}%
x_{q}x_{\bar{q}}\left(  x_{\bar{q}}+x_{g}\delta_{-s_{g}\lambda_{q}}%
-\delta_{s\lambda_{q}}\right)  \right\}  -\left(  q\leftrightarrow\bar
{q}\right) \, ,\nonumber\\
\label{F1eT}%
\label{resF2tildeT}
&&\hspace{-.8cm}\tilde{F}_{2}\left(  p_{q},p_{\bar{q}},p_{g},k,\vec{z}_{1},\vec{z}_{2}\right)
^{\alpha}\varepsilon_{T\alpha}=-4g\,\theta(p_{g}^{+}-\sigma)\,\delta(k^{+}%
-p_{g}^{+}-p_{q}^{+}-p_{_{\bar{q}}}^{+})\frac{e^{-i\vec{p}_{q} \cdot \vec{z}%
_{1}-i\vec{p}_{_{\bar{q}}} \cdot \vec{z}_{2}}}{\sqrt{2p_{g}^{+}}}\delta_{\lambda
_{q},-\lambda_{\bar{q}}}%
\nonumber \\
&&\times  
\frac{\left(  \delta
_{\lambda_{\bar{q}}s}-x_{q}\right)  \left(  \delta_{-s_{g}\lambda_{q}}%
x_{g}+x_{\bar{q}}\right)}{x_{\bar{q}} \, x_g}  
\frac{\vec{P}_{\bar{q}} \cdot %
\vec{\varepsilon}_{g}^{\,\,\ast}}{\vec{P}_{\bar{q}}^2}
\frac{\vec{z}_{12} \cdot \vec{\varepsilon}_{T}}{\vec{z}_{12}^2} 
\, Q \, Z_{122}
K_{1}(QZ_{122})e^{-i\vec{p}_{g} \cdot \vec{z}_{2}%
}-\left(  q\leftrightarrow\bar{q}\right)  \,. 
\end{eqnarray}
We denote 
$ F_{2}\left(  p_{q},p_{\bar{q}},p_{g},z_{0},\vec{z}_{1},\vec{z}_{2}\right)^{\alpha}\!=\!\tilde{F}_{2}\left(  p_{q},p_{\bar{q}},p_{g},z_{0},\vec{z}_{1}%
 ,\vec{z}_{2}\right)  ^{\alpha}\!+\!\int d\vec{z}_{3}\,F_{1}\left(  p_{q},p_{\bar{q}%
 },p_{g},z_{0},\vec{z}_{1},\vec{z}_{2},\vec{z}_{3}\right)  ^{\alpha}\!.$

\section{2- and 3-gluon approximation}

Let us notice that the dipole operator $\mathbf{U}_{ij}$ involves terms at least of order $g^2$. Hence for only two or three exchanged gluons one can neglect the quadrupole term in the amplitude $M^{\alpha}$ which results in the simpler expression
\begin{eqnarray}
\label{M3gBis}
&& M^{\alpha} \overset{\mathrm{g^3}}{=}   \frac{1}{2}\int d\vec{z}_{1}d\vec{z}%
_{2} \mathbf{U}_{12}  \left[ \left(  N_{c}^{2}-1\right)
\tilde{F}_{2}\left(  \vec{z}_{1},\vec{z}%
_{2}\right) ^{\alpha} \right.
\nonumber \\
&& \left.
+ \int d\vec{z}_{3} \left\{  N_{c}^{2}F_{1}\left(
\vec{z}_{1},\vec{z}_{3},\vec{z}_{2}\right)^{\alpha} 
+N_{c}^{2}F_{1}\left(  \vec{z}_{3},\vec{z}%
_{2},\vec{z}_{1}\right)  ^{\alpha} -  F_{1}\left(  \vec{z}_{1},\vec{z}_{2},\vec{z}_{3}\right)  ^{\alpha} \right\} \right].
\end{eqnarray}
For $\vec{p}_q=\vec{p}_g=\vec{p}_{\bar{q}}=\vec{0}$, those integrals can be performed analytically. Otherwise they can be expressed as a simple convergent integral over $[0,1]$ that can be performed numerically for any future phenomenological study. 

\section{Towards the next-to-leading-order corrections}

\begin{figure}[h]
\centerline{\includegraphics[scale=0.85]{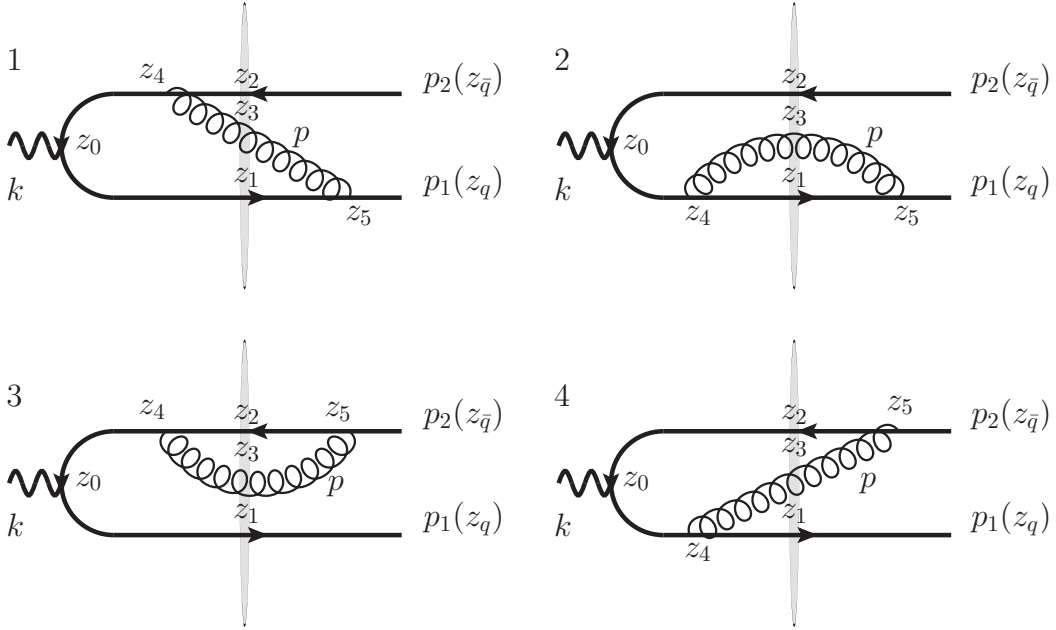}}
\caption{Diagrams contributing to virtual corrections in which the radiated gluon interacts with the shock wave.}
\label{nloSW}
\end{figure}

The calculation of virtual corrections to the $\gamma^* \to q \bar{q}$ involves two kinds of contributions.
\begin{figure}[h]
\centerline{\includegraphics[scale=0.85]{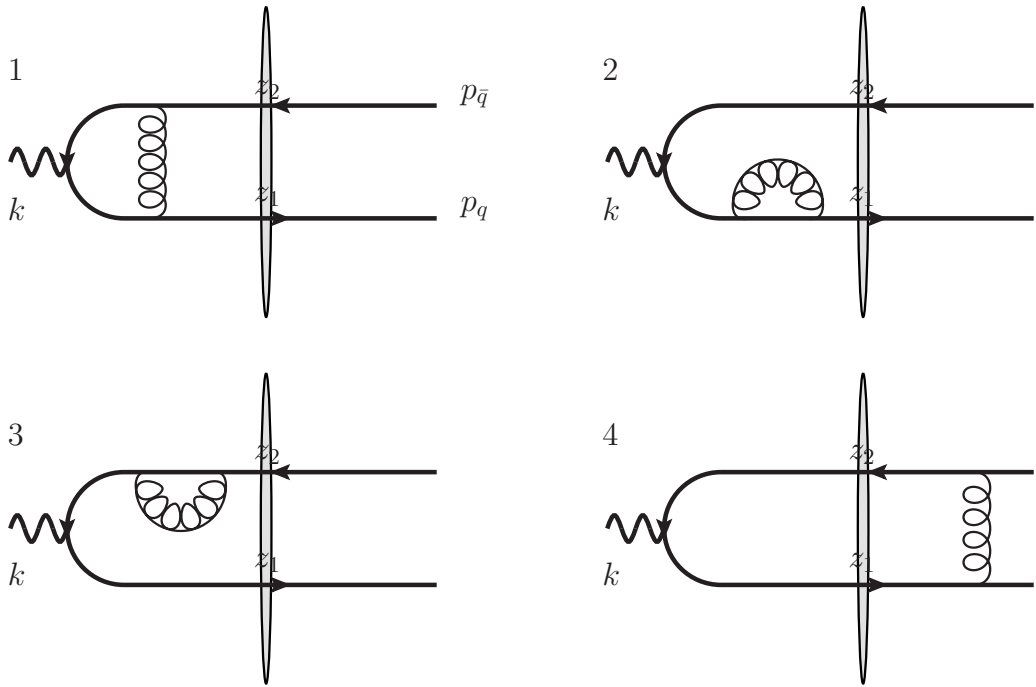}}
\caption{Diagrams contributing to virtual corrections in which the radiated gluon doesn't cross the shock wave.}
\label{nlo}
\end{figure}
\noindent
The diagrams contributing to virtual corrections in which the radiated gluon does not cross the shock wave are shown in Fig.~\ref{nlo},
whereas the diagrams in which the radiated gluon interacts with  the shock wave are illustrated in the Fig.~\ref{nloSW}.

This calculation is much more complicated than the computation of the real contribution to the $\gamma^* \to q \bar{q} g$ impact factor discussed above. Although it only involves one-loop diagrams, the complications arise due to the presence of many different scales. Indeed, our aim is to obtain results in the general kinematics where the virtuality of incoming photon, the $t-$channel momentum transfer and the invariant mass $M^2$ of the diffractive two-jet state are arbitrary. Additionally, this impact factor  is a function of the virtuality of $t-$channel exchanged gluons. This work is in progress.
We here present the matrix element corresponding to the first three diagrams of Fig.~\ref{nlo}. We work in 
dimensional regularization for the transverse momentum space, i.e. $d=D-2 =2+ 2\,\epsilon\,,$ and introduce the regularization scale $\mu$, and the related dimensionless scale  $\tilde{\mu}^2=\mu^2/Q^2\,.$
Denoting $p_{ij} \equiv p_i-p_j\,,$ we introduce $p_\perp=p_{q1\perp}\,,$  $\vec{p}^{\,2}=-p_\perp^2$ and
$w=\vec{p}^{\,2}/Q^2\,.$ For simplicity, we write $x=x_q.$
We get for the case of a longitudinally polarized photon 
\begin{eqnarray}
&&\hspace{-1cm}T_{fi}|_{\epsilon_{\alpha}=n_{2\alpha}}=-i g^2\frac{N_{c}^{2}-1}{2N_{c}%
}tr(U(p_{1\bot})U^{\dag}(-p_{2\bot}))\delta(p_{q1\bot}-p_{\gamma\bot}%
+p_{\bar{q}2\bot})\delta(p_{q}^{+}-p_{\gamma}^{+}+p_{\bar{q}}^{+})\theta
(p_{q}^{+})\theta(p_{\bar{q}}^{+})
\nonumber \\
&\times&\frac{\Gamma(1-\epsilon)}{\left(
16\pi^{3}\right)  ^{1+\epsilon}}\frac{1}{\sqrt{2p_{\gamma}^{+}}\sqrt
{2p_{q}^{+}}\sqrt{2p_{\bar{q}}^{+}}}
\frac{x(1-x)p_{\gamma}^{+}{}\overline{u}_{p_{q}}\gamma^{+}v_{p_{\bar{q}%
}}}{x(1-x)Q^{2}+\vec{p}{}^{\,\,2}} 
\nonumber \\
&\times& \left(  \left(  2\ln\left(  \frac
{(1-x)x}{\alpha^{2}}\right)  -3\right)  \left(  \ln\left(  \frac{\left(
w-x^{2}+x\right)  ^{2}}{(1-x)x\tilde{\mu}^{2}}\right)  +\frac{1}{\epsilon
}\right)  +\ln^{2}\left(  \frac{x}{1-x}\right)  -\frac{\pi^{2}}{3}+6\right)  \,. 
\end{eqnarray}
Expanding the photon momentum in the Sudakov basis (\ref{Sudakov-basis}) as
\begin{equation}
p_\gamma = p_\gamma^+ \, n_1 - \frac{Q^2}{2 p_\gamma^+} \, n_2
\end{equation}
one can explicitly check the electromagnetic gauge invariance for this group of diagrams
since
\begin{equation}
T_{fi}|_{\epsilon_{\alpha}=n_{1\alpha}}=\frac{Q^{2}}{2p_{\gamma}^{+2}}%
T_{fi}|_{\epsilon_{\alpha}=n_{2\alpha}}\,.
\end{equation}
Similarly, for the case of a transversally polarized photon, one gets
\begin{eqnarray}
&& \hspace{-1cm}T_{fi}|_{transverse}=-i g^2 \frac{N_{c}^{2}-1}{2N_{c}}tr(U(p_{1\bot})U^{\dag
}(-p_{2\bot}))\delta(p_{q1\bot}-p_{\gamma\bot}+p_{\bar{q}2\bot})\delta
(p_{q}^{+}-p_{\gamma}^{+}+p_{\bar{q}}^{+})\theta(p_{q}^{+})\theta(p_{\bar{q}%
}^{+})
\nonumber \\
&\times&
\frac{\Gamma(1-\epsilon)}{\left(  16\pi^{3}\right)  ^{1+\epsilon}%
}\frac{\epsilon_{i}}{\sqrt{2p_{\gamma}^{+}}\sqrt{2p_{q}^{+}}\sqrt{2p_{\bar{q}%
}^{+}}}  \frac{-\left(  \frac{1}{2}\overline{u}_{p_{q}}[\gamma^{i}\hat{p}_{\bot
}]\gamma^{+}v_{p_{\bar{q}}}+(2x-1)p^{i}\overline{u}_{p_{q}}\gamma
^{+}v_{p_{\bar{q}}}\right)  }{2(x(1-x)Q^{2}+\vec{p}{}^{\,\,2})}%
\nonumber \\
&\times&\left[  \left(  2\ln\left(  \frac{(1-x)x}{\alpha^{2}}\right)  -3\right)
\left(  \ln\left(  \frac{w-x^{2}+x}{\tilde{\mu}^{2}}\right)  +\frac
{(1-x)x\ln\left(  \frac{(1-x)x}{w-x^{2}+x}\right)  }{w}+\frac{1}{\epsilon
}\right)  
\right. \nonumber \\
&& \left.
+\,\ln^{2}\left(  \frac{x}{1-x}\right)  -\frac{\pi^{2}}{3}+6\right]  .
\end{eqnarray}

\section{Conclusion}

The measurement of dijet production in DDIS was recently performed~\cite{Aaron:2011mp}, and a precise comparison of 
dijet versus triple-jet production, which has not been  performed yet at HERA~\cite{Adloff:2000qi}, would be very useful to get a deeper understanding of the QCD mechanism underlying diffraction. Recent investigations of the azimuthal distribution of dijets in diffractive photoproduction performed by ZEUS~\cite{Guzik:2014iba} show sign of a possible need for a 2-gluon exchange model, which is part of the shock-wave mechanism. Our calculation could be used for phenomenological studies of those experimental results.
A similar and very complementary study could be performed at LHC with UPC events. A full quantitative first principle analysis of this will be possible only after completing our program of  computing virtual corrections to the $\gamma^* \rightarrow q\bar{q}$ impact factor, as discussed above.

Diffractive open charm production was measured at HERA~\cite{Aktas:2006up} 
and studied in the large $M$ limit based on the direct coupling between a Pomeron and a $q \bar{q}$ or a $q\bar{q}g$ state, with massive quarks~\cite{Bartels:2002ri}. Such a program could also be performed at LHC, 
again based on UPCs and on
the extension of the above mentioned impact factors to the case of a massive quark. It is the subject of presently ongoing research. Beyond jets, this could be further extended 
to $J/\Psi$ mesons, which are copiously produced at LHC.

\ack

We thank Ian Balitsky, Cyrille Marquet and St\'ephane Munier for discussions.

A. V. G. acknowledges support of president scholarship 171.2015.2,
RFBR grant 13-02-01023, Dynasty foundation, Metchnikov grants and University Paris Sud. He 
is also
grateful to LPT Orsay for hospitality  while part of the
presented work was being done. R. B. thanks RFBR for financial support
via grant 15-32-50219.
This work was partially supported by the PEPS-PTI PHENODIFF,
the PRC0731 DIFF-QCD, the Polish Grant NCN No. DEC-2011/01/B/ST2/03915, the ANR PARTONS (ANR-12-MONU-0008-01), the COPIN-IN2P3 Agreement and the Th\'eorie-LHC France Initiative.

\end{document}